%% file: main.tex
\documentclass[a4paper,11pt]{article}
\usepackage{aaskaiid}
\usepackage{orcidlink}
\usepackage{pdflscape} % for landscape page
\usepackage[usenames,dvipsnames,table,svgnames]{xcolor}
\definecolor{carmine}{rgb}{0.59, 0.0, 0.09} % line color

\newcommand{\eg}[0]{$\textnormal{e.g. }$}
\newcommand{\ie}[0]{$\textnormal{i.e. }$}
\newcommand{\tn}[1]{\textnormal{#1}}
\newcommand{\sub}[1]{_{\textnormal{#1}}}

\usepackage{xspace}
\newcommand{\HI}{\textsc{Hi}\xspace}
\title{\HI Simulations for Cosmology with the SKA Observatory}
\ShortTitle{\HI Simulations}

\author[1,2]{Tommaso Ronconi \orcidlink{0000-0002-3515-6801}}
\ShortName{Ronconi, De Lucia et al.} % shortened name list for header 
\author[3,7]{Gabriella De Lucia \orcidlink{0000-0002-6220-9104}}
\author[4,19]{Marta Spinelli \orcidlink{0000-0003-0148-3254}}
\author[5]{Pascal Hitz \orcidlink{0000-0003-0479-5405}}
\author[3,7]{Fabio Fontanot \orcidlink{0000-0003-4744-0188}}
\author[6,7]{Francesco Sinigaglia \orcidlink{0000-0002-0639-8043}}
\author[8]{Lizhi Xie \orcidlink{0000-0003-3864-068X}}
\author[18]{Robert M. Yates \orcidlink{0000-0001-9320-4958}}
\author[20]{Jos\'e Luis Bernal \orcidlink{0000-0002-0961-4653}}
\author[9]{Anna Bonaldi \orcidlink{0000-0003-4787-2888}}
\author[10,11,12]{Stefano Camera \orcidlink{0000-0003-3399-3574}}
\author[3]{Isabella P. Carucci \orcidlink{0000-0001-5287-0065}}
\author[13,14]{Steven Cunnington \orcidlink{0000-0001-6594-107X}}
\author[21,22]{Francisco-Shu Kitaura \orcidlink{0000-0002-9994-759X}}
\author[5]{Jo\"{e}l Mayor \orcidlink{0009-0008-6499-8013}}
\author[15]{Rajesh Mondal \orcidlink{0000-0001-7728-3756}}
\author[10,11]{Giulia Piccirilli \orcidlink{0000-0002-3341-1872}}
\author[16,17]{Digvijay Wadekar \orcidlink{0000-0002-2544-7533}}

\affiliation[1]{INAF -- Institute of Radioastronomy (IRA), Via Gobetti 101, 40129 Bologna, Italy}
\affiliation[2]{Scuola Internazionale Superiore di Studi Avanzati, Via Bonomea 265, 34136 Trieste, Italy}
\emailAdd{tommaso.ronconi@inaf.it}
\affiliation[3]{INAF -- Osservatorio Astronomico di Trieste, Via Tiepolo 11, 34131 Trieste, Italy}
\emailAdd{gabriella.delucia@inaf.it}
\affiliation[4]{Observatoire de la C\^ote d’Azur, Laboratoire J-L Lagrange, Boulevard de l'Observatoire, Nice, France}
\affiliation[5]{ETH Zurich -- Institute for Particle Physics and Astrophysics, 8093 Zurich, Switzerland}
\affiliation[6]{University of Geneva, 24 rue du Général-Dufour, 1211 Genève, Switzerland}
\affiliation[7]{IFPU: Institute for Fundamental Physics of the Universe, Via Beirut, 2, 34151 Trieste, Italy}
\affiliation[8]{Tianjin Normal University, Binshuixidao 393, 300387, Tianjin, China}
\affiliation[9]{SKA Observatory, Jodrell Bank, Lower Whitington, Macclesfield, SK11 9FT, UK}
\affiliation[10]{Dipartimento di Fisica, Università degli Studi di Torino, Via P. Giuria 1, 10125 Torino, Italy}
\affiliation[11]{INFN, Sezione di Torino, Via P. Giuria 1, 10125 Torino, Italy}
\affiliation[12]{INAF, Osservatorio Astrofisico di Torino, Via Osservatorio 20, 10025 Pino Torinese, Italy}
\affiliation[13]{Institute of Cosmology \& Gravitation, University of Portsmouth, Dennis Sciama Building, Portsmouth, PO1 3FX, UK}
\affiliation[14]{ Jodrell Bank Centre for Astrophysics, Department of Physics and Astronomy, The University of Manchester, Manchester M13 9PL, UK}
\affiliation[15]{Department of Physics, National Institute of Technology Calicut, Calicut 673601, Kerala, India}
\affiliation[16]{Department of Physics and Astronomy, Johns Hopkins University, 3400 N. Charles Street, Baltimore, Maryland, 21218, USA}
\affiliation[17]{School of Natural Sciences, Institute for Advanced Study, 1 Einstein Drive, Princeton, NJ 08540, USA}
\affiliation[18]{Centre for Astrophysics Research, University of Hertfordshire, Hatfield, AL10 9AB, UK}
\affiliation[19]{Department of Physics \& Astronomy, University of the Western Cape, Cape Town 7535, South Africa}
\affiliation[20]{Instituto de F\'isica de Cantabria (IFCA), CSIC-Univ. de Cantabria, Avda. de los Castros s/n, E-39005 Santander, Spain}
\affiliation[21]{Instituto de Astrof\'{\i}sica de Canarias, Vía Láctea s/n, E-38205, La Laguna, Tenerife, Spain}  
\affiliation[22]{Departamento de Astrof\'{\i}sica, Universidad de La Laguna, E-38206, La Laguna, Tenerife, Spain}

\abstract{We present a comparative overview of state-of-the-art methods for modelling the distribution of neutral hydrogen (\HI) in the post-reionization Universe, developed in preparation for upcoming SKAO cosmological surveys. Our aim is to assess how different physical and empirical assumptions reflect into predictions for key observables such as the cosmic \HI density, the \HI mass function, and the \HI-halo mass relation. 
We consider both: (i) semi-analytical approaches that self-consistently evolve baryonic components within dark matter merger trees through physically motivated prescriptions and (ii) empirical schemes tailored to different observables and based on fast approximations designed for large ensemble studies. 
By comparing the predictions from the different methods considered, we find overall consistency in integrated quantities such as $\Omega_{\mathrm{HI}}$, yet systematic differences in the detailed shape and scatter of the \HI--halo mass relation and its redshift evolution. Semi-analytical models offer physically grounded predictions but depend on assumed prescriptions, while empirical methods provide flexibility and computational efficiency at the expense of robustness in extrapolated regions of the parameter space.
The increasing number of \HI measurements from SKA precursors and pathfinders (including surveys with MeerKAT, ASKAP, and FAST) will provide critical observational constraints to refine and calibrate current simulation methodologies. In turn, increasingly realistic \HI simulations play a key role in interpreting these data, guiding survey design and analysis strategies, in preparation for the advent of SKAO data.}

\begin{document}
\maketitle

%\section{Introduction}
%Reference .

\section{Introduction}

Modelling of neutral hydrogen (\HI) across cosmic time is a key requirement for forthcoming large-scale surveys with the SKA Observatory \citep[SKAO;][]{braun2019anticipatedperformancesquarekilometre} and its precursors, that will exploit \HI-based observables to address both astrophysical and cosmological questions.
On large scales, the distribution and evolution of \HI trace the underlying matter density field \citep[e.g.][]{Chang2010,Masui2013,Padmanabhan2015,Spinelli_etal_2020}, while on galactic scales, the \HI reservoir fuels star formation and regulates the interplay between accretion, feedback, and the enrichment of the interstellar and circumgalactic media \citep[e.g.][]{Furlanetto2006,PritchardLoeb2010,Crain+17,Diemer2019,ReinaCampos2022}.
By connecting the dark and baryonic components of the Universe, \HI serves as a bridge between cosmology and galaxy evolution, providing insight into structure formation, gas cycling, and the efficiency of galaxy growth.
In the post-reionization era, \HI is thus a powerful tracer of both the expansion history of the Universe, the physics of baryons and their interplay with the dark sector \citep[e.g.][]{blyth2015aska,santos2015aska}.

The SKAO observational capabilities will allow a variety of survey strategies. For cosmological studies, 
these include \HI\ galaxy surveys, exploiting resolved detections of the 21-cm emission line to perform cosmological analyses via, for example, redshift-space distortions and baryon acoustic oscillations \citep{Mayor01.2026.SKA, Nasirudin01.2026.SKA}, and \HI\ intensity mapping (IM), where the integrated 21-cm signal from unresolved sources is measured to probe large-scale modes with high efficiency \citep{Wolz01.2026.SKA, Mazumder01.2026.SKA}. 
Each approach requires accurate theoretical predictions and realistic mock catalogues to design optimal survey strategies, validate analysis pipelines, and interpret forthcoming observations.

The generation of \HI mock data relies on a broad spectrum of modelling techniques, ranging from fully hydrodynamical simulations \citep[e.g.][]{Crain+15,Crain+17,VillaescusaNavarro2018, Stevens2019} to semi-analytical galaxy formation models and empirical or statistical prescriptions calibrated against observations. 
While hydrodynamical simulations capture the detailed interplay between baryons and dark matter, they are still computationally expensive and limited in volume \citep{Lagos01.2026.SKA}. Semi-analytical and empirical methods, by contrast, offer the flexibility and scalability required for cosmological applications, enabling the production of large-volume realisations and light-cones suitable for the design and interpretation of SKAO surveys.

This Chapter provides an overview and comparison of state-of-the-art techniques developed to model the \HI content of the post-reionization Universe, from galaxy-level catalogues to large-scale intensity maps. 
These methods are relevant to many of the scientific topics discussed in other chapters of this volume, including cosmological inference from 21-cm intensity mapping \citep{Wolz01.2026.SKA, Majumdar01.2026.SKA}, as well as studies of \HI-selected galaxies for cosmology and astrophysics \citep{Mayor01.2026.SKA, Nasirudin01.2026.SKA}. 
By presenting a diverse set of approaches, this Chapter aims to give an overview of assumptions and limits of available methodologies for the design of future \HI surveys and the interpretation of their data in preparation for the SKAO era. 
In fact, \HI simulations are essential for forecasting survey performance and for validating analysis pipelines. 
They provide controlled environments in which to test the impact of instrumental effects, foreground removal, and systematic uncertainties on recovered cosmological signals, as discussed in \citet{Spinelli01.2026.SKA}. 
Mock realisations are also crucial for estimating covariance matrices and quantifying cosmic variance in higher-order statistics \citep{Majumdar01.2026.SKA}. 
As the community advances toward the first SKAO observations, the interplay between realistic \HI simulations, mock survey generation, and data-analysis methodologies will be central to transforming 21-cm observations into robust cosmological constraints.

This Chapter is organised into three main Sections. 
In Sec.~\ref{sec:methods}, we introduce a representative, though not exhaustive, sample of methodologies developed to simulate neutral hydrogen, encompassing both \HI galaxy catalogues and approaches tailored for \HI intensity mapping (IM). In Sec.~\ref{sec:comparison}, we present a comparative analysis of the selected methodologies based on the simulations described above. Finally, in Sect.~\ref{sec:discussion}, we conclude by outlining the main results of our comparison, highlighting the general trends, limitations, and strengths of the different methods and providing an outlook on future developments informed by upcoming observations.

\section{Methods}\label{sec:methods}

In recent years, substantial progress has been made in modelling the interstellar medium (ISM) within cosmological hydrodynamical simulations, with several suites of zoom-in or small-volume simulations now able to follow gas cooling down to temperatures of $\lesssim 10$~K \citep{Agertz+20, reina-campos2023, feldmann2023, schaye2025}. 
Such treatments, however, require extremely high mass and spatial resolution, and are therefore unfeasible for large cosmological volumes. 
To overcome these limitations, a range of complementary methods have been developed, that aim to predict the large-scale distribution of \HI\ while avoiding the high computational cost of full hydrodynamical calculations.

In this Section we focus on five representative methods, selected to illustrate the heterogeneous strategies currently available for simulating the large-scale \HI distribution. 
These include both semi-analytical and empirical frameworks and differ in their complexity, predictive power, and calibration strategies.

The sample of methods presented here is by no way exhaustive. 
Several additional approaches have been proposed in the last $\sim15$~years. 
For example, the semi-numerical pipeline of \citet{mondal2015, mondal2016, mondal2018} combines a parallelized particle-mesh $N$-body simulation\footnote{\url{https://github.com/rajeshmondal18/N-body}} to generate the dark matter density field with a friends-of-friends halo finder\footnote{\url{https://github.com/rajeshmondal18/FoF-Halo-finder}} to identify collapsed structures. 
Neutral hydrogen is then assigned to haloes following the \HI--halo mass relation proposed by \citet{bagla2010}. 
The resulting \HI\ distribution is used to construct redshift-space 21-cm light-cones for observability forecasts. 
Empirical \HI\-host halo relations can also be applied to lightcones with semi-empirical astrophysical properties obtained from e.g., UniverseMachine~\citep{Behroozi:2019kql}, coherent with simulated emission from other tracers of the LSS, as done in~\citet{Sato-Polito:2022wiq}.
Other groups have proposed techniques that exploit machine learning to associate an \HI distribution to the output of dark matter only simulations. 
For example, \citet{Wad20_ML} introduced an augmented \HI--halo mass relation that extends the traditional $M_{\rm HI}$--$M_h$ scaling by incorporating secondary halo properties $\{i_h\}$ such as environment, concentration, spin, and formation epoch. 
The relation $M_{\rm HI}=f_{\rm ML}(M_h,\{i_h\})$ is learnt using random forest and symbolic regression techniques trained on hydrodynamical simulations \citep{VillaescusaNavarro2018}. Therefore, the approach explicitly accounts for assembly-bias that is traditionally neglected in standard empirical approaches.  
Subsequently, \citet{Wad21_HInet} presented a convolutional neural network (CNN) model that maps the dark matter density field, $\rho_{\rm m}(\mathbf{x})$, to the corresponding \HI\ density, $\rho_{\rm HI}(\mathbf{x})$, producing high-resolution \HI\ boxes from coarse-grained dark matter inputs and enabling fast generation of realistic large-scale \HI\ fields.

In certain applications, multiple ($\mathcal{O}(100)$) realisations of \HI\ simulations are required—such as to estimate covariances in clustering measurements or to reconstruct signal loss due to background cleaning in \HI intensity mapping experiments \citep[e.g.][]{Switzer:2015ria, Cunnington:2023jpq, MeerKLASS:2024ypg}. 
A fast mock-generation process is essential in these cases and many of the previously mentioned techniques, while accurate, are still too computationally demanding to this aim. A more direct approach for generating density fields consists in producing realisations from a prescribed power spectrum. 
Such fields, by construction, reproduce their input power spectrum with variance determined by the simulation volume (i.e.\ the sample variance). 
The simplest implementation involves generating Gaussian random realisations of the input $P(k)$; however, a log-normal transformation can be applied, at negligible additional cost, to obtain a field whose probability distribution function more closely resembles that of the cosmic matter density \citep{Shin:2017cwu}. 
Such simulations can be realised, for example, by adapting to the \HI case the Generator for Large Scale Structure (GLASS; \citealp{Tessore_2023}), that generates highly detailed and realistic galaxy survey simulations at high angular resolution while efficiently optimising computational resources.
Additional realism can be incorporated by, for instance, Poisson-sampling galaxies or adding velocity-field realisations \citep[e.g.][]{Agrawal:2017khv, Murray2018}. Then, the line luminosity of each galaxy can be sampled from an input luminosity function to include a realistic distribution of intensities in the resulting map, as done in SIMPLE~\citep{LujanNiemeyer:2023vfz}.

In the following sub-sections, we describe the specific methods used in this work, that form the basis for the comparison presented in Sec.~\ref{sec:comparison}.

\subsection{Semi-Analytical Models}\label{sec:methods:sam}

In the semi-analytical approach, the evolution of the baryonic components of dark matter haloes is modelled using simple but physically motivated prescriptions that are coupled to a description of the assembly history of dark matter haloes, which in the jargon is referred to as `merger tree'. The latter can be either based on analytical approaches (e.g. extended Press-Schechter formalism, eventually with corrections based on results from high-resolution dark matter simulations - see for example \citealt{Parkinson_etal_2008}) or extracted from numerical simulations \citep[e.g.][]{Tweed_etal_2009}. By adopting this approach, it is possible to express the galaxy formation process essentially through a set of (coupled) differential equations that govern the exchanges of mass and energy between the different baryonic components of a galaxy. Given our limited understanding of the physical processes at play, the description involves a number of free parameters (and parametrizations) that are typically tuned to reproduce a given subsample of observational measurements in the local Universe.

The approach is very flexible and is characterized by limited computational costs, which gives access to a large dynamic range in mass and spatial resolution in large computational volumes allowing an efficient exploration of the parameter space and of the influence of specific physical mechanisms. However, assumptions need to be made to compensate for the lack of an explicit treatment of the gas dynamics, and for the lack of information about the spatial distribution of baryonic matter.  

The two models introduced below are state-of-the-art models that find roots in the original model developed around the `Munich' group \citep[][and references therein]{Springel_etal_2001,DeLucia_etal_2004,Croton_etal_2006,DeLucia_and_Blaizot_2007}. 
They are both coupled to substructure based merger trees, i.e. the resolution of the backbone simulation is high enough to resolve the cores of haloes accreted onto larger structures (up to some limit cosmic time, depending on the resolution, in time and mass, of the simulation). 
Below, we give an overview of the two models, with references to the original work where all details can be found. 

\subsubsection{GAEA (sam1)} \label{sec:method:GAEA}
The GAlaxy Evolution and Assembly (GAEA) model is based on the original model introduced in \citep{DeLucia_and_Blaizot_2007} but it has been significantly updated since then. In particular, the most recent rendition of the model \cite{DeLucia_etal_2024} includes: (i) a detailed chemical enrichment scheme that accounts for the finite stellar lifetimes and the associated recycling of metals, energy and gas \citep{DeLucia_etal_2014}; (ii) an updated stellar feedback scheme that is partially based on results from high-resolution hydrodynamical simulations \citep{Hirschmann_etal_2016}; (iii) an updated treatment for gas accretion onto supermassive black holes and the associated feedback \citep{Fontanot_etal_2020} and an updated treatment for the stripping of hot gas associated with infalling satellites and cold gas stripping from satellite galaxies \citep{Xie_etal_2020}. The model also includes a treatment for the partition of cold gas in its atomic and molecular gas component, and a star formation law that is based on the molecular gas \citep{Xie_etal_2017}. The model version adopted in this Chapter uses empirical relations based on the hydrostatic pressure of the disc \citep[][previous work shows that alternative prescriptions lead to very similar predictions at the galaxy masses and redshift ranges considered here]{blitz2006}.  

GAEA is tuned to reproduce the observed galaxy stellar mass function and the AGN luminosity function up to $z\sim 3$, and the \HI and H$_2$ galaxy mass functions measured in the local Universe. It has been shown to successfully reproduce a number of observational measurements over a wide range of cosmic epochs. Relevant for this paper: model predictions are in quite good agreement with observational measurements of scaling relations between the amount of gas in different phases and galaxy stellar mass in the local Universe, both for central and satellite galaxies \citep[][see also \citealt{Zoldan_etal_2017}]{Xie_etal_2020}. In addition, the model reproduces the observed clustering strength of galaxies selected on the basis of their \HI mass \citep{Fontanot_etal_2025}. In \citet{Spinelli_etal_2020} the authors presented a comprehensive analysis of model predictions for the \HI content of dark matter haloes, based on a previous version of our model and extending up to $z\sim 5$.

In the default run used for this Chapter, GAEA is coupled to high-resolution cosmological N-body simulations \citep{Fontanot_etal_2025}. In \citet{Cammelli_etal_2025} the model has been coupled to merger trees extracted from PINOCCHIO \citep[PINpointing Orbit Crossing Collapsed HIerarchical Objects, ][]{Monaco_2016}, which allows access to higher resolution cosmological boxes and the construction of several independent mocks with limited computational time. Finally, dedicated software has been developed to associate an \HI emission line with each simulated galaxy in post-processing \citep{mayor2026}.
Developed for GAEA but also tested on L-Galaxies, this allows realistic modelling, validated against observational data, of line parametrisation (e.g. width) statistics, scaling relations (e.g. Tully-Fisher) and \HI galaxy survey selection systematics.
This software, as well as outputs from GAEA,  are used in other Chapters in this Book \citep[\ie][]{Nasirudin01.2026.SKA,Mayor01.2026.SKA} to derive predictions for \HI galaxy survey number counts, redshift distributions, and \HI galaxy bias as functions of instrument sensitivities. 

\paragraph{HIP-POP:}\label{sec:method:hiphop}

Building upon the baryonic prescriptions derived from the \textsc{GAEA} semi-analytical model, the \HI\ Probe Populator (\texttt{HIP-POP}\footnote{\url{https://github.com/spinemart/HIP-POP}}) provides a fast, approximate method to populate large-scale dark matter halo catalogues with neutral hydrogen for intensity mapping applications, where the limited angular resolution does not require modelling of  individual galaxies.

The key ingredient is the $M_{\rm HI}-M_{\rm halo}$ relation, modelled following \citet{Spinelli_etal_2020} as:
\begin{equation}
    M_{\mathrm{HI}}(M_h)=M_h \left[a_1 \left(\frac{M_h}{10^{10}}\right)^{\beta} e^{-\left(\frac{M_h}{M_{\mathrm{break}}}\right)^\alpha}+a_2\right] e^{-\left(\frac{M_{\mathrm{min}}}{M_h}\right)^{0.5}},
    \label{eq:M_HI}
\end{equation}
where $M_\text{HI}$ and $M_\text{halo}$ are the hydrogen and halo mass, while $a_1$, $\beta$, $\alpha$, $M_{\rm break}$, $a_2$, and $M_{\rm min}$ are free parameters. 
Their redshift evolution is derived from fits to \textsc{GAEA} outputs across multiple snapshots, using a Gaussian likelihood and the \textsc{multinest} sampler \citep{Feroz2009} through an MPI-enabled Python wrapper \citep{Zwart2016}. 
Parameter values and scatter are interpolated in redshift using spline fits up to $z=1$.

This semi-empirical framework allows for the rapid generation of large-volume, full-sky \HI mock maps consistent with the adopted semi-analytical galaxy formation model.

\subsubsection{L-Galaxies (sam2)}\label{sec:method:yates}

As mentioned above, L-Galaxies \citep{Croton_etal_2006,Guo+11,Henriques+15,Henriques+20,Yates+24} has roots back to similar models on which GAEA is based. However, the models have developed completely independently over the past $\sim 15$ years and differ significantly in terms of the physical prescriptions adopted. 
The L-Galaxies version adopted in this Chapter \citep{Yates+24} includes the following advanced prescriptions: (i) a detailed chemical enrichment scheme including binary stars \citep{Yates+13,Yates+24} that tracks a full complement of 118 chemical elements produced from 16 different enrichment sources; (ii) an explicit treatment for dust production and destruction \citep{Vijayan+19,Yates+24}; (iii) a modelling for radially resolved gas and stellar discs \citep{Fu+13}. 

As for GAEA, L-Galaxies includes an explicit treatment for the partition of cold gas in \HI and H$_2$. 
The model version considered in this Chapter follows the metallicity- and density-dependent formalism from \citet{McKee&Krumholz10}. 
This formalism returns an H$_{2}$ fraction of $0$ at densities below $\sim{}30\ M_\odot\tn{pc}^{-2}$ and $1$ at densities above $\sim{}100\ M_\odot\tn{pc}^{-2}$. 
The L-Galaxies partitioning prescription returns a good match to the $z=0$ \HI mass function (by construction) and to the $M\sub{HI}/M_{*}$ fraction \citep{Henriques+20}. 
However, the cosmic \HI and H$_{2}$ mass densities are lower than inferred from observations at $z\sim{}3-4$, as it is the case for many other semi-analytical models, including GAEA. 
Likely, this offset is due to the lack of cold gas modelling at the interface between the  ISM and CGM \citep{Yates+21b}.

The L-Galaxies source code is fully open source and available on \href{https://github.com/LGalaxiesPublicRelease/LGalaxies2020_PublicRepository/tree/Yates2023}{GitHub}, with detailed installation instructions provided on the \href{https://lgalaxiespublicrelease.github.io/index.html}{L-Galaxies website}. Outputs from various L-Galaxies versions are also available for download from the \href{https://gavo.mpa-garching.mpg.de/Millennium/}{Millennium database} (free registration required), with selected outputs from the latest version of the model also available via \href{https://zenodo.org/records/10019005}{Zenodo}.

\subsection{Empirical Models}\label{sec:empirical}

An alternative approach to link dark matter haloes with the expected properties of \HI galaxies, or more generally with the large-scale distribution of neutral hydrogen, is to employ simulation methodologies that are, to some extent, directly tuned to observational data of the baryonic properties they aim to reproduce. 
The primary goal of these methods is not to explicitly model the physical processes responsible for the emergence of the observables, but rather to rely on empirical assumptions and approximations in order to generate outputs that reproduce the same statistical properties of the selected galaxy population observed at a given stage of its evolution.

These models are designed to bypass the uncertainties associated with sub-grid physics in simulations of galaxy formation and evolution. 
Instead, they rely on predefined recipes either to sample populations of mock objects with given observable properties or to assign such properties to a background simulated DM field. 
While, on the one hand, these models are computationally convenient, on the other, their predictive power is inherently limited. 
The reason for this is twofold: 
first, their outputs are only as reliable as the observational datasets used for calibration; 
and second, when the model predictions extend beyond the range over which they were tuned, they rely on extrapolations that may not accurately represent the real Universe.

In most cases and in the framework relevant for this Chapter, the main goal of these approaches is to deliver a credible statistical description of clustering as a function of redshift. This is achieved by assuming a specific \HI mass-halo mass relation ($M_{\rm HI}-M_{\rm halo}$ or \HI HM). The latter can be conditioned to environmental properties derived from physically motivated arguments or based on results from a parent hydrodynamical simulation. 
With additional assumptions, the techniques can be extended to provide explicit \HI distributions for galaxies populating dark matter haloes. In this Chapter, we consider predictions from three different empirical methods. In the following subsections, we provide a summary of their specific implementations.

\subsubsection{T-RECS + SCAM (emp1)}
\label{sec:method:ronconi}

This is a publicly available pipeline that simulates radio sources by overlaying them on a halo/sub-halo distribution derived from cosmological N-body simulations. The pipeline enables the generation of full-sky light-cones of \HI\ galaxies along with their continuum counterparts. Full details are given in \cite{Ronconi2026}. The main components of the method can be summarized as follows: 
\begin{itemize}
    \item N-body simulations provide a structural framework to identify dark matter haloes and sub-haloes, that host galaxies (both the line-emitting and continuum populations). Specifically, the pipeline employed in this Chapter uses a full-sky light-cone created by tiling snapshots from the DEMNUni simulation set \citep{Carbone2016,Parimbelli2022} up to redshift $z\sim5$, specifically leveraging the high-resolution run \citep[HR-DEMNUni, e.g.][]{Hernandez-Molinero2024}.
    \item Catalogues of continuum and line-emitting galaxies are generated using T-RECS \citep[Tiered Extragalactic Radio Continuum Simulations,][]{Bonaldi2018,Bonaldi2023}, based on empirical measurements. The mock populations are sampled independently to reproduce observed statistics, such as the \HI mass function (up to redshift $z\sim0.5$) and the variation of \HI flux with mass, as well as the source size and morphology distributions. 
    \item To ensure that the synthetic galaxies are accurately coupled with the dark matter distribution, the pipeline applies the SCAM (Sub-halo Clustering and Abundance Matching) method \citep{Guo2016}, implemented in the Python library SCAMPy \citep{Ronconi2020}. 
    SCAM follows a two-step approach: first, two independent Halo Occupation Distribution (HOD) models are used to select the sub-haloes that host the two populations considered. 
    These are tuned to match the clustering data of \HI galaxies, star-forming galaxies and active galactic nuclei independently.
    As a second step, a Sub-halo Abundance Matching algorithm is used to connect synthetic T-RECS sources with the DM haloes and sub-haloes selected in the first step. 
\end{itemize}
The final mock catalogues provide detailed information for \HI line emission, continuum sources, and their dark matter host halos. These products include comprehensive source catalogues (up to redshift $z<0.5$ for the \HI galaxies and up to $z=5$ for continuum sources) and pixelated maps. All codes described above are publicly available. 
Given that the independent populations are associated to the same light-cone, cross-catalogues can be extracted as well.

In the first release \citep{Ronconi2026}, two catalogues are made available:
\begin{enumerate}
    \item The first catalogue is tuned to reproduce a realistic spatial distribution of sources by exploiting existing measurements of the two-point correlation function. 
    The parameters of the HOD model are therefore calibrated by maximizing a Gaussian likelihood that simultaneously accounts for both the one-point and two-point statistics, namely the observed number density of sources and their clustering signal:
    \begin{equation}
        \label{eq:chi2_1-2pt_scampy}
        \mathcal{L}(\boldsymbol{\theta}) = -\dfrac{1}{2}\left[\chi^2_{n_g}(\boldsymbol{\theta}) + \chi^2_{\xi_r}(\boldsymbol{\theta})\right],
    \end{equation}
    where $\boldsymbol{\theta}$ denotes the free parameters of the HOD model, $n_g$ is the spatial number density of sources, and $\xi_r$ represents the scale-dependent two-point correlation function in real space. 
    The observational datasets employed to constrain the HODs of the two main source populations in T-RECS are: the three-dimensional correlation function of \HI-selected galaxies from \citet{Martin2012}, and the angular correlation function of continuum-selected sources (SFGs and AGN) from \citet{Hale2018}.

    \item A second, deeper catalogue is generated by neglecting the clustering term in Eq.~\eqref{eq:chi2_1-2pt_scampy}, thereby omitting an explicit modelling of the two-point statistics. 
    In this case, the two independent HOD models are calibrated solely to reproduce the expected redshift evolution of the source number density, as predicted for the SKA-A\!A4 survey specifications \citep{Hartley2023}.
\end{enumerate}

\subsubsection{HI Halo Model (emp2)}
\label{sec:method:hitz}
\citet{Hitz2025} present a fast method for simulating the cosmological distribution of post-reionization \HI based on a halo model approach, that is tailored for forecasting and forward modelling of intensity mapping experiments. 
The method can be summarised as follows:
\begin{itemize}
    \item Past Light Cone of Dark Matter Halos: the method employs the approximate but fast PINOCCHIO code \citep{Monaco2002} to simulate the past light cone of dark matter halos. PINOCCHIO is based on Lagrangian Perturbation Theory together with the Extended Press \& Schechter formalism for non-spherical collapse \citep{Bond1991, Monaco1995, Monaco1997} to efficiently generate halo catalogues  while maintaining clustering properties at the 10\% level accuracy compared to N-body simulations. \cite{Hitz2025} describes a simulation suitable for the survey configuration of the HIRAX (Hydrogen Intensity and Real-time Analysis eXperiment, \citealt{HIRAX2016, HIRAX2022}) radio interferometer array. Specifically, a light cone covering declinations between $-15^{\circ}$ and $-45^{\circ}$ in the frequency range 700--800 MHz (corresponding to redshifts 0.77--1.03) is generated. PINOCCHIO is run with parameters allowing to achieve a halo mass resolution of $4.3 \times 10^9\, \text{M}_{\odot}$, leading to an estimated missing \HI fraction of 2--3\% over the considered frequency range.
    
    \item \HI-Halo Mass Relation: the \HI mass is subsequently assigned to dark matter halos based on their masses using a \HI-halo mass relation  constrained by observational data \citep{Padmanabhan2017}:
    \begin{equation}
    M_\text{HI}(M_\text{halo},z) = \alpha f_\text{H,c} M_\text{halo} \biggl (\frac{M_\text{halo}}{10^{11} \, h^{-1} \text{M}_{\odot}} \biggr)^{\beta} \exp \biggl[- \Bigl( \frac{v_\text{c,0}}{v_\text{c}(M,z)} \Bigr)^3 \biggr],
    \label{eq:Pascal MHI-M relation}
    \end{equation}
    where $f_\text{H,c} = (1-Y_\text{p})\Omega_\text{b} / \Omega_\text{m}$ is the cosmic fraction of \HI with respect to all matter, $M_\text{HI}$ and $M_\text{halo}$ are the \HI and halo mass, respectively, $v_\text{c}$ is the virial velocity of a host halo, while $\alpha$, $\beta$ and $v_\text{c, 0}$ are free fitting parameters. 
    In their work, the authors obtain the best-fit values $\alpha = 0.09$, $\beta = -0.58$, and $\log v_\text{c,0} = 1.56$.
    The virial velocity as a function of halo mass \citep{Barnes2014} in the exponential introduces a cut-off in \HI mass at low halo masses, reflecting that smaller haloes cannot effectively self-shield \HI from ionizing radiation.
    
    \item Map Generation: in the final step, the past light cone is divided into 20 frequency bins, each with a bandwidth of 5 MHz, to generate pixelated \texttt{HEALPix} maps with an \textit{NSIDE} parameter of 512. Maps are generated for dark matter and \HI halo masses, halo counts, and \HI brightness temperature. The latter is computed using the relation from \cite{Bull_2015}.
\end{itemize}
The simulation has been validated by comparing the mass function, mass density, three-dimensional power spectrum, and angular power spectrum of the maps with predictions from an analytical dark matter and \HI halo model, implemented in PyCosmo \citep{Refregier2018, Tarsitano2021, Moser2022, Hitz2025}, a Python package for theoretical cosmological predictions. 
The simulated data set, including the light-cone halo catalogue, snapshot halo catalogues ($1 \text{h}^{-3} \text{Gpc}^3$ box at $z=0.8,0.9,1$), as well as the \texttt{HEALPix} maps, is available at \url{https://cosmology.ethz.ch/research/software-lab/cosmological-neutral-hydrogen-simulation.html}.

\subsubsection{Environment-dependent HIHM relation (emp3)}
\label{sec:method:sinigaglia}

This method produces full-sky mock catalogues of \HI sources built directly on a light-cone, with a smooth redshift evolution extending up to $z \simeq 3.8$. 
It combines approximate dark matter simulations with a model that links dark matter haloes to their expected neutral hydrogen content. 
The process can be summarised in three main steps:
\begin{itemize}
    \item Dark matter light-cone: the dark matter distribution is generated with the \texttt{WebON} code \citep{Kitaura2026}, that implements the \textit{Augmented Lagrangian Perturbation Theory} \citep[\texttt{ALPT};][]{Kitaura2013}.  
    In this framework, the growth of cosmic structures is followed by displacing particles from their initial to final positions using precomputed displacement fields.  
    The light-cone is built by interpolating outputs at different redshifts and applying a correction based on the Zel’dovich approximation \citep{AnguloWhite2010}.  
    This provides a fast, approximated, but accurate way to reproduce the large-scale structure expected from gravitational evolution.

    \item Populating the light-cone with haloes: the dark matter field is then populated with dark matter haloes using the \texttt{Hicobian} bias model \citep{ColomaNadal2024}, that connects the matter distribution to the expected halo abundance through a non-linear and stochastic relation \citep[\eg,][]{Kitaura2014}.  
    The model accounts for the diversity of cosmic environments (such as voids, filaments, sheets, and knots) by classifying each region according to its local gravitational field \citep{Hahn2007, HeavensPeacock1988, McDonaldRoy2009, Sinigaglia2021, Sinigaglia2022, Kitaura2022, ColomaNadal2024}.  
   The model parameters are calibrated using the \textsc{AbacusSummit} suite of $N$-body simulations \citep{Maksimova2021, Garrison2019, Garrison2021}, exploring a broad range of cosmological scenarios around the {Planck18} cosmology \citep{Planck2018}.

    \item Assigning halo and \HI properties:  once the number of haloes is defined in each region, the method assigns them positions, velocities, and masses.  
    Halo positions are initially matched to dark matter particles and then slightly adjusted to reproduce the observed small-scale clustering \citep{ForeroSanchez2024}.  
    Redshift-space distortions are modelled following \cite{Kaiser1987}, including both coherent large-scale motions and random velocities within haloes \citep[e.g.,][]{Kitaura2012a, Kitaura2012b, Hess2013, Kitaura2014, Kitaura2016, Bos2019, Sinigaglia2022, Sinigaglia2024a, Sinigaglia2024b}.  
    Halo masses are drawn from analytic mass functions, while the amount of neutral hydrogen is assigned using the HIHM relation measured in hydrodynamical simulations from \cite{VillaescusaNavarro2018}.  
    This relation is extended to include environmental effects, allowing the \HI content to depend not only on halo mass but also on the surrounding large-scale structure \citep{Sinigaglia2026}.  
    All parameters are redshift-dependent and are interpolated between simulation snapshots to produce a continuous, redshift-evolving \HI catalogue.
\end{itemize}

\section{Comparison}\label{sec:comparison}

Having introduced the main methodologies developed by the different teams, we now turn to a comparison between some of their relevant outputs. 
The goal of this exercise is not to establish a ranking among models, but rather to illustrate how different methodological assumptions and calibration strategies reflect in the predicted statistical properties of neutral hydrogen across cosmic time. 
This comparison also highlights the relative strengths and limitations of the available approaches.
A summary of the models contributing to this comparison is provided below:
\begin{itemize}
    \item[\textbf{sam1:}] the \textsc{GAEA} semi-analytical model described in Sec.~\ref{sec:method:GAEA} (results shown as green diamonds);
    \item[\textbf{sam2:}] the \textsc{L-Galaxies} semi-analytical model described in Sec.~\ref{sec:method:yates} (results shown as light-blue plus markers);
    \item[\textbf{emp1:}] divided into emp1.1 (lavender upward triangles) and emp1.2 (golden squares), corresponding to the two \textsc{T-RECS+SCAMPy} catalogues described in Sec.~\ref{sec:method:ronconi};
    \item[\textbf{emp2:}] the \HI halo-model introduced in Sec.~\ref{sec:method:hitz} (blue downward triangles);
    \item[\textbf{emp3:}] the environment-dependent HIHM-relation method from Sec.~\ref{sec:method:sinigaglia} (orange crosses).
\end{itemize}

The comparison presented in this Section is neither homogeneous nor exhaustive, as the models differ in several fundamental aspects:
\begin{itemize}
    \item \textbf{Baryonification strategy:} the sam1 and sam2 models follow a semi-analytical approach, while the remaining models are empirical, each relying on distinct prescriptions for connecting dark matter haloes and/or sub-haloes with baryonic observables.
    \item \textbf{Cosmological parameters:} these depend on the underlying dark-matter simulations used to generate the mock data. Specifically, all methods have been, or can be, run with different cosmologies.  
    In this Section, we use simulation runs consistent with a \textit{flat}-$\Lambda$CDM cosmology, which, though, vary minimally in the specific parameters chosen.
    The complete set of cosmological parameters employed in each case is listed in Tab.~\ref{tab:cosmologies}.
    \item \textbf{Simulated properties:} sam1, sam2 and emp1 provide galaxy-level information, whereas emp2 and emp3 only provide the total neutral hydrogen content associated with host haloes.
    \item \textbf{Geometry:} sam1 and sam2 output data in comoving simulation boxes at discrete redshifts, while the other three models produce full-sky light-cones. For both emp2 and emp3, results can be extracted both from a light-cone and from the underlying simulation boxes from which it was constructed.
    \item \textbf{Available redshifts:} the redshift coverage varies among the models, being related to the depth of the light-cones for the empirical models, and determined by the granularity of the snapshots for the semi-analytic models.
\end{itemize}

\begin{table}
    \centering
    \captionof{table}{Fiducial Flat-$\Lambda$CDM cosmological parameters used for the present manuscript. Note that these values are not necessarily the only possible set of parameters available for the listed method, only the parameters used for the measurements presented in Sec.~\ref{sec:comparison}. \textit{References:} (1) \citet{Baugh_etal_2019}; (2) \citet{Springel_etal_2005}; (3) \citet{Boylan-Kolchin_etal_2009}; (4) \citet{Carbone2016}; (5) \citet{Parimbelli2022}; (6) \citet{Hernandez-Molinero2024}; (7) \citet{Monaco2002}; (8) \citet{Kitaura2026}.}
    \label{tab:cosmologies}
    \input{tab/cosmologies}    
\end{table}

As a consequence of this heterogeneity, a direct one-to-one comparison across all statistics is not always feasible. 
In the following subsections, we specify on a case-by-case basis which models are compared, under what conditions, and for which physical quantities.

\subsection{HI cosmic density}\label{sec:omhi}

\begin{figure}
    \centering
    \includegraphics[width=0.95\linewidth]{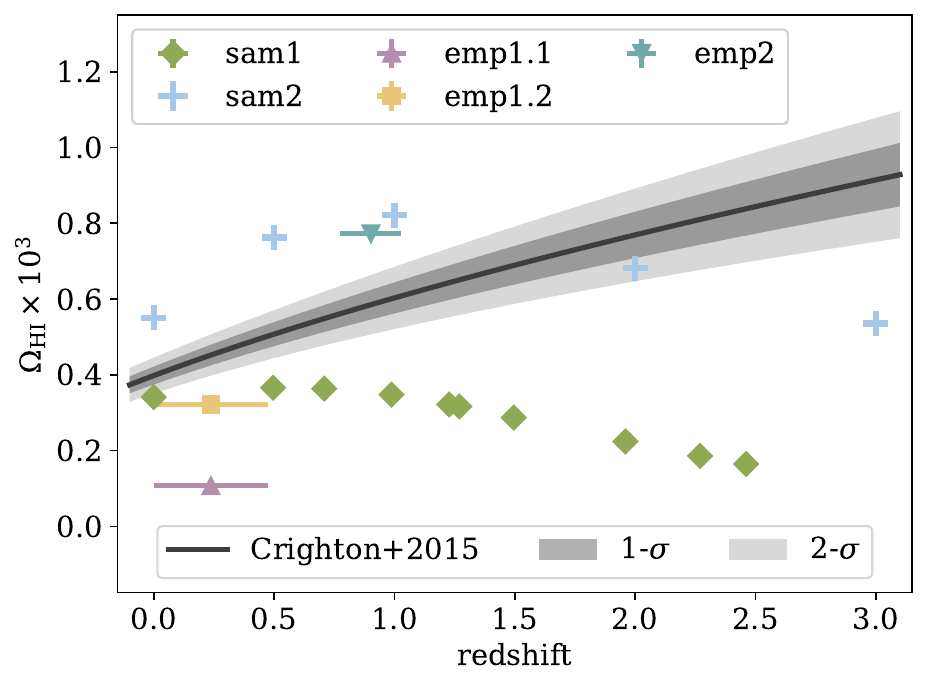}
    \caption{Measurements of $\Omega_\text{HI}$ at different redshifts from 3 methods out of the 4 listed in the Chapter. For comparison, we also mark the empirical relation found in \cite{Crighton2015} that represents a fit to observational data.}
    \label{fig:OmHI}
\end{figure}

Here we compare the \HI cosmic density derived for the different simulations considered in this Chapter.
This has been obtained by summing up all the hydrogen contained either in the box or along the light-cone, and dividing the resulting volume density by the critical density parameter for the corresponding cosmology:
\begin{equation}
    \label{eq:OmHI}
    \Omega_\text{HI}(z) = \dfrac{1}{V(z) \rho_\text{crit, 0}}\sum_i M_\text{HI}^i
\end{equation}
where $V(z)$ is the volume of each redshift slice considered in the case of light-cones, while it corresponds to the volume of the simulation box for sam1 and sam2  (only snapshots are available for these).
For each different method, the sum runs over all simulated systems in the catalogue: galaxies for sam1, sam2 and emp1 and host haloes for emp2.
The critical density at redshift zero, $\rho_\text{crit, 0}$ is computed consistently with the cosmological parameters of the background simulation.

In Fig.~\ref{fig:OmHI} the results of the sam1, sam2, emp1 and emp2 methods are reported with different coloured markers and error-bars.
The redshift dimension error corresponds to the thickness of the redshift slice considered and is thus not present for sam1 and sam2 whose boxes are associated to discrete redshift values.
The $\Omega_\text{HI}$-dimension error was computed by bootstrap resampling both the boxes and the redshift slices.
Since it varies between $0.01\%\text{--}0.1\%$, for all the models and redshifts considered, it is not visible in the Figure.

As a term of comparison, we also show (solid black line with shaded grey regions marking the 1$\sigma$ and 2$\sigma$ confidence intervals) the empirical relation for the evolution of the hydrogen density parameter derived by \citet[][hereafter C15]{Crighton2015}. 
The authors compiled measurements from independent studies, converted them to a common cosmology, and computed average values of the density parameter from redshift $z=0$ up to $z=5$. 
The solid line with its confidence regions in Fig.~\ref{fig:OmHI} represents a power-law of the form $\Omega_\mathrm{HI} = A (1+z)^\gamma$, fitted to the dataset considered in C15. 
The confidence regions were obtained by Gaussian sampling around the mean and standard deviation of the best-fitting parameters: $A = (4.00\pm0.24) \times 10^{-4}$ and $\gamma = 0.60\pm0.05$. 
We note that \citet[][hereafter PH20]{PH2020} later fitted the same functional form to an updated compilation of measurements, finding consistent values, namely $A = (4.6\pm0.2) \times 10^{-4}$ and $\gamma = 0.57\pm0.04$. 
In this work, we report only the C15 relation, as most of the models considered here were calibrated on data predating 2020.
We also note that both C15 and PH20 rely on \HI emission measurements at low redshift, and on absorption-based estimates to constrain the neutral hydrogen abundance at earlier cosmic times.

Both the {sam1} and {sam2} models are calibrated to reproduce the \HI galaxy mass function at redshift $z\sim0$, and predict its evolution at higher redshift through physically motivated prescriptions. 
As shown in the figure, the green diamond symbols corresponding to predictions from {sam1} are systematically below the empirical relation with deviations increasing towards higher redshift. 
As discussed in \citet{Spinelli_etal_2020} and \citet{Yates+21b}, this behaviour is at least in part expected, since the \HI budget predicted by SAMs only accounts for the gaseous content of galaxies and does not include gas in filaments and in the circum-galactic medium (CGM). The {sam2} model qualitatively reproduces the observed decreasing trend below $z\sim1$, but also deviates from observational measurements at higher redshifts. \citet{Yates+21b} showed that this is the case even when accounting for sub-resolution haloes and for missing \HI in the IGM \citep[see also][]{DiGioia_etal_2020}.

The emp1 model also simulates galaxies, but uses an empirical approach. 
The emp1.1 run shown as a pink triangle in Fig.~\ref{fig:OmHI} matches the clustering properties of the \HI galaxy population analysed in \cite{Martin2012} which is calibrated on a sub-sample of the $\alpha.40$ catalogue lacking a large fraction of low-\HI-mass sources.
For this reason, the total neutral hydrogen budget available from its mock sources is only $\sim35\%$ of the observational constraint.
On the other hand, the deeper emp1.2 run of the same model reaches the same $\Omega_\text{HI}$ value measured on all the other datasets available in the same redshift bins. 
Indeed, this run is explicitly tuned to reproduce the expected number counts of objects in the local Universe, instead of their 2-point statistics.

For the {emp2} method we have access to a light-cone within a narrow redshift interval, $0.77 < z < 1.03$, as it has been originally developed in the context of preparatory studies for HIRAX. Their adopted configuration and mass resolution allows the recovery of more than 97 per cent of the expected \HI density. Fig.~\ref{fig:OmHI} shows that the cosmic density from emp2 is very close to that predicted by sam2 at $z=1$  and actually above the fitting formula by \citet{Crighton2015}. This difference arises from the used empirical \HI-halo mass relation adopted from \cite{Padmanabhan2017}, which is characterised by a cut-off at considerably low masses (\ie $\log M_\text{HI}/M_\odot\lesssim 10\text{--}11$), due to the inefficiency of \HI self-shielding (see Sec.~\ref{sec:method:hitz}).

\subsection{\HI mass function}\label{sec:himf}

We measure here the \HI mass function of galaxies from sam1, sam2 and emp1.
We do not include the other models in this comparison because they do not allow to directly extract information on the \HI mass function of galaxies from their outputs: even though the emp2 model is based on methodology \citep{Padmanabhan2017} that allows to extract the \HI mass function from the \HI content of haloes, this information is not available for single galaxies. Similarly, emp3 does not natively provide information on individual galaxies but only on the total host halo hydrogen content.

While sams provide a fixed redshift binning, emp1 is defined continuously in redshift, and is upper limited at $z \leq 0.5$.
We limited this comparison to 3 redshift slices: the 2 redshift values for which information are available from all the three simulation methods considered here, and redshift $z=1$ where only results from sam1 and sam2 are available.
For the emp1 case, we consider measurement for thin slices of the light-cone, spanning the intervals $z\in [0.0, 0.015)$ and $z\in[0.475, 0.525)$.

\begin{figure}
    \centering
    \includegraphics[width=0.95\linewidth]{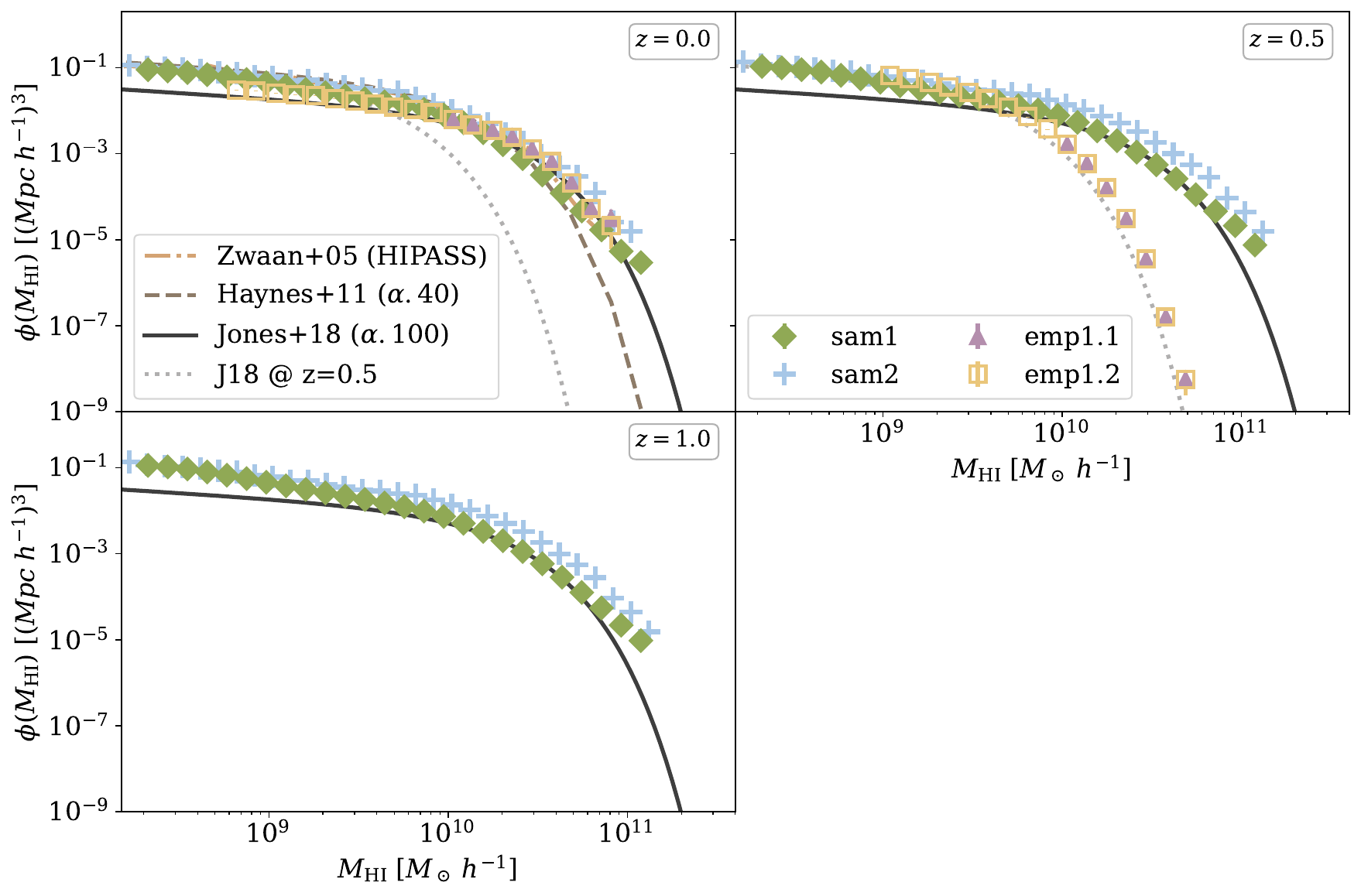}
    \caption{HI mass functions at redshift $z=0$ (left panel) and $z=0.5$ (right panel). Measurements from sam1 (green diamonds), sam2 (light-blue plus symbols), emp1.1 (lavender triangles) and emp1.2 (empty golden squares) are shown along with analytical models fitted to data. The models shown in the upper left panel represent fit to different survey data: HIPASS \citep[bronze dashed-dotted line,][]{Zwaan+05}, ALFALFA-$40\%$ \citep[gray dashed line,][]{Haynes+11} and ALFALFA-$100\%$ \citep[black solid line][]{Jones+18}.
    On the 2 upper panels we also show an extrapolation of the redshift-dependent model fitted by \cite{Bonaldi2023} to data from redshift $z\sim0$ \citep{Jones+18} to $z\sim0.3$ \citep{Bera2022,Paul2023}.}
    \label{fig:HIMF}
\end{figure}
The results of this comparison are shown in the three panels of Fig.~\ref{fig:HIMF}, for the three redshift values considered.
Green diamonds and light-blue plus symbols are for sam1 and sam2 respectively, while the other two markers correspond to the 2 light-cones of the emp1 model: lavender triangles markers are for emp1.1 and empty golden squares for emp1.2.

In the upper left panel, we over-plot different fit to  survey data from the literature, as a term of comparison.
The bronze dashed-dotted line shows the HIPASS mass function \citep{Zwaan+05}; the gray dashed line corresponds to the results obtained with $40\%$ of the ALFALFA survey \citep{Haynes+11}, while the black solid line shows the most up-to-date estimate from the full ALFALFA survey \citep[J18 hereafter]{Jones+18}.
Note that all these Schechter-like mass function fits have been obtained mainly with data at redshift $z=0$. 
In the upper- and lower-left panels, we only show the most recent J18 fit.
Along with these fairly established fitting functions, we also mark the behaviour, extrapolated up to $z=0.5$, using the redshift-dependent model presented in \cite{Bonaldi2023}, tuned to reproduce the limited amount of data available for redshift $z>0$, namely the Schechter function obtained in \cite{Bera2022} at $z=0.35$ and the one obtained by indirectly deriving the HIMF from the \HI power spectrum at redshifts $z=0.32$ and $z=0.44$ \citep{Paul2023}.
This HIMF model is the one used in \cite{Bonaldi2023} to sample sources for the T-RECS mock-catalogues and is thus representative of the simulated dataset of emp1. 

At redshift $z=0$, all methods considered reproduce well the observational data by construction, with an excess of galaxies with \HI masses below $\sim 10^{10}$ for the sam1 and sam2 methods.
The main difference between the 4 simulations considered is on the \HI-mass resolution limit. 
The sam1 and sam2 models can be considered complete down to  $M_\text{HI} \gtrsim 10^{8.5} M_\odot/h$ and  $M_\text{HI} > 10^7 M_\odot/h$, respectively  (see original papers).
The emp1 model is limited by construction to higher mass values due to the necessity of reproducing clustering data (i.e., roughly, only sub-haloes with higher masses are selected as galaxy hosts), with emp1.1 complete up to $M_\text{HI} \gtrsim 10^{10} M_\odot/h$ and emp1.2 $M_\text{HI} \gtrsim 10^{9} M_\odot/h$.
In all simulations, the high mass end of the \HI mass function is limited by the limited size of the simulated volume.

An important difference between the models considered in this section is visible in the upper right panel of Fig.~\ref{fig:HIMF}, for the \HI mass function at $z=0.5$.
While the sam1 and sam2 models predict very little evolution of the \HI mass function up to $z\sim 1$, the emp1 model is tuned to reproduce a strong evolution of the \HI mass function between $z=0$ and $z=0.5$ (gray dotted line). 
As mentioned above, little observational information is currently available beyond $z=0$. 
However, the situation is rapidly evolving with on-going surveys, such as WALLABY \citep{wallaby2020}, MIGHTEE-HI \citep{mightee2021} and LADUMA \citep{laduma2016}, and will dramatically change when the SKAO will become operational.

\subsection{\HI to halo mass relation}\label{sec:hihm}

The \HI to halo mass (HIHM) relation provides insight on the connection between dark matter halos and their baryonic content.

Instead of performing a direct measurement, for {sam1} we employed {HIP-POP}\footnote{See the dedicated paragraph in Section~\ref{sec:method:hiphop}. Note that, HIP-HOP has been trained on data from a version of GAEA pre-dating the one used to measure the statistics presented in this Section. Nevertheless, the difference is negligible with respect to the accuracy of our comparative analysis.}, an approximate method calibrated on {sam1} statistics. 
It therefore encodes the same information as {sam1} within the validity range of the fits, namely $\log M_\text{halo} \in [9.5, 13.5]$ for halo mass and $z \in [0, 1]$ for redshift. 
The {emp2} method can construct light-cones by adopting the HIHM relation form fitted by \citet{Padmanabhan2017}; we therefore use this analytical relation directly for comparison, since the current implementation of {emp2} does not include scatter around its functional form and thus coincides exactly with it. 
For {emp1}, given the limited depth of the light-cone, we obtained results only at low redshift.\footnote{We show only the {emp1.2} result, as {emp1.1} contains the same information at lower resolution.}

For methods producing galaxy catalogues (\ie sam1, sam2 and emp1), the HIHM relation is computed by summing the \HI masses of all galaxies belonging to a given halo.
This approach is prone to underestimates of the total \HI content, as neither the faintest galaxies nor the diffuse, unbound neutral hydrogen are included in the total (as already commented in Sec.~\ref{sec:omhi}). 

\begin{figure}
    \centering
    \includegraphics[width=0.95\linewidth]{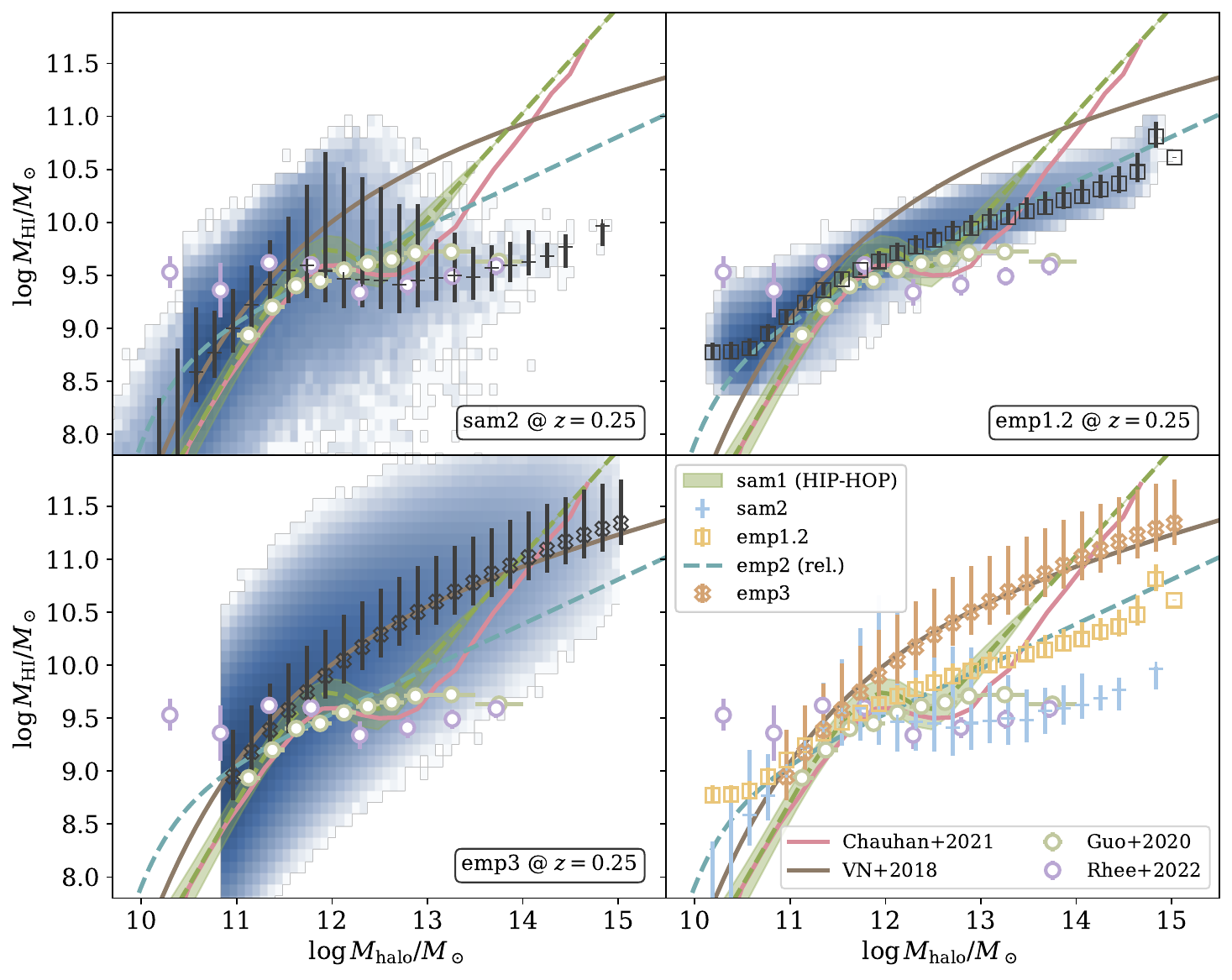}
    \caption{Comparison of the \HI–halo mass (HIHM) relation at $z=0.25$ obtained from all the models considered in this work.
    The upper panels and lower left panel show the percent abundance of haloes in the $(M_{\mathrm{halo}}, M_{\mathrm{HI}})$ plane
    for sam2 (upper left), emp1 (upper right) and emp3 (lower left).
    Blue shaded regions indicate the density of simulated haloes, while empty black squares with error bars
    mark the mean and standard deviation of $\log M_{\mathrm{HI}}$ in bins of $\log M_{\mathrm{halo}}$.
    Circles with error bars correspond to observational measurements of central galaxies in clusters 
    \citep{Guo2020, Rhee2023}, and solid lines represent predictions from hydrodynamical simulations 
    \citep{VillaescusaNavarro2018, Chauhan2021}.
    The lower right panel summarizes all models and reference datasets for direct comparison.}
    \label{fig:HIHMz025}
\end{figure}
In Fig.~\ref{fig:HIHMz025} we show, in all panels but the one on the lower right, the distribution of \HI mass per halo of given mass as obtained from the catalogues available for sam2 (upper left panel), emp1 (upper right panel) and emp3 (lower left panel). The distributions are shown as blue shaded regions while the empty black squares with vertical error bars indicate the mean and standard deviation of $\log M_{\mathrm{HI}}$ in bins of $\log M_{\mathrm{halo}}$ for each of these distributions.  
These summary statistics are all shown together in the lower right panel, with the same colouring convention adopted in this Section, for comparison.
We also show the results of sam1 and emp2 in all the panels with dashed lines.

As a reference, we also show results from the literature in all the panels: circles with error bars correspond to observational measurements of central galaxies in clusters \citep{Guo2020, Rhee2023}, while solid lines show predictions from hydrodynamical simulations \citep{VillaescusaNavarro2018, Chauhan2021}. 
The lower right panel provides a direct comparison between all datasets and methods.
The semi-analytical models (both sam2 shown in the upper left panel, and sam1 marked by the \texttt{HIP-HOP} fit) show the same dip present in \cite{Chauhan2021}, that is a consequence of AGN feedback. We can also see the transition from the higher-resolution catalogue (run on Millennium-II) to the lower-resolution catalogue (run on Millennium) at log$(M_{\rm halo}/M)\sim{}10.5$ in the density binning for sam2 in the upper left panel.

The emp1 method (upper right panel) exhibits a normalisation and shape of the relation comparable with observational constraints but no dip. 
Over the mass range considered, it is also close to the relation based on emp2, marked by the blue dashed line (all panels). 
However, contrary to emp2, the distribution obtained from emp1 is not the result of an explicit tuning but rather an indirect outcome of the sub-halo abundance-matching scheme.
The emp3 model (lower left panel) was tuned to reproduce the relation from \cite{VillaescusaNavarro2018} and thus, it matches it closely across the full mass range. 

\begin{figure}
    \centering
    \includegraphics[width=0.95\linewidth]{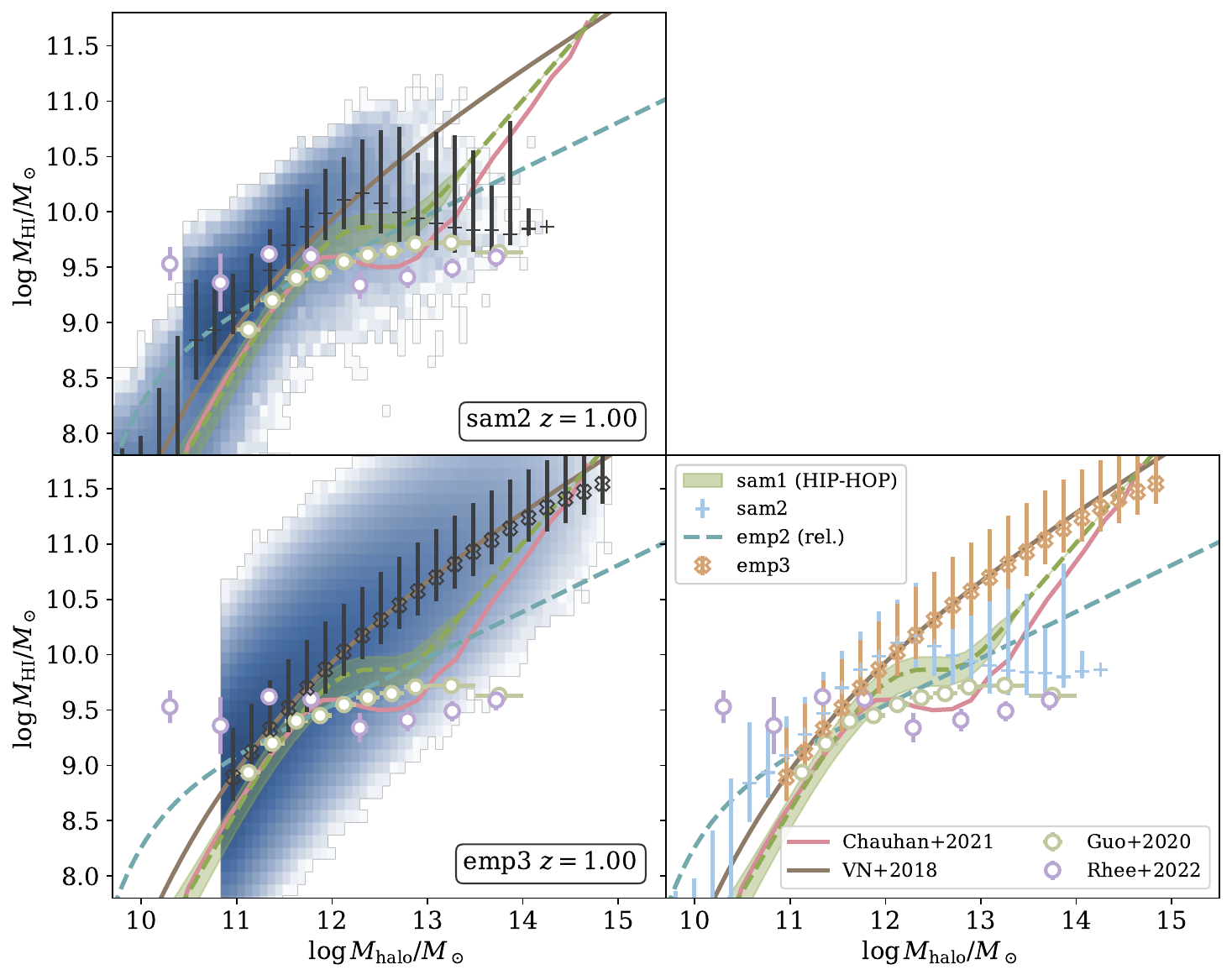}
    \caption{Same as Fig.~\ref{fig:HIHMz025}, but for redshift $z=1$.
    The upper and lower left panels show results from the sam2 and emp3 models, respectively.}
    \label{fig:HIHMz100}
\end{figure}
In Fig.~\ref{fig:HIHMz100} we present the same comparison as in Fig.~\ref{fig:HIHMz025}, at $z \approx 1$. 
This redshift value is out of the emp1 light-cone and, thus, the corresponding upper right panel is not present.
The overall behaviour of the different models considered is the same described for Fig.~\ref{fig:HIHMz025}, i.e. the differences between the methods and observational constraints do not appear to change significantly up to $z\sim 1$. 
We note that there is a fairly good agreement across simulation methodologies and observational constraints at intermediate halo masses, while the results are significantly more scattered at large halo masses where, due to the limited volumes of current \HI surveys, constraints from observational datasets are scarce. Additional observational measurements (both in the radio and in the optical wavelength regime) will enable a better characterization of the HIHM relation, both for the most massive haloes and at halo masses $\sim 10^{12}\, M_{\odot}$, where observational uncertainties on halo mass affect the strength of the dip that some models predict. 

\subsection{Marked power spectrum}\label{sec:pk}

The marked power spectrum is a weighted generalisation of the standard two-point clustering statistic, designed to probe correlations between large-scale structure and a local property (the \emph{mark}) of the underlying field \citep{sheth2005,white2016,satpathy2019}. 
Given a tracer density field $n(\mathbf{x})$ and an associated scalar mark $m(\mathbf{x})$, the marked density contrast is defined as
\begin{equation}
    \delta_m(\mathbf{x}) = \frac{m(\mathbf{x})\,n(\mathbf{x})}{\langle m\,n \rangle} - 1,
\end{equation}
where $\langle \cdot \rangle$ denotes the spatial mean. 
The \emph{marked power spectrum} is then the Fourier-space two-point statistic of this marked field:
\begin{equation}
    P_m(k) = \langle \tilde{\delta}_m(\mathbf{k})\,\tilde{\delta}_m^{*}(\mathbf{k}) \rangle,
\end{equation}
that reduces to the standard matter power spectrum $P(k)$ when the mark is $m(\mathbf{x}) = 1$ everywhere.

The $P_m(k)$ is sensitive to how clustering depends on the chosen mark, which can encode non-linear or environmental effects.
Common choices include marks that depend on the local density, halo mass, or other internal properties such as the stellar mass, the gas fraction, or specifically in our case, the \HI mass content of the objects studied.
In this way, the marked spectrum traces the coupling between large-scale modes and small-scale physics, extending beyond what is captured by $P(k)$ alone.

\begin{figure}
    \centering
    \includegraphics[width=0.95\linewidth]{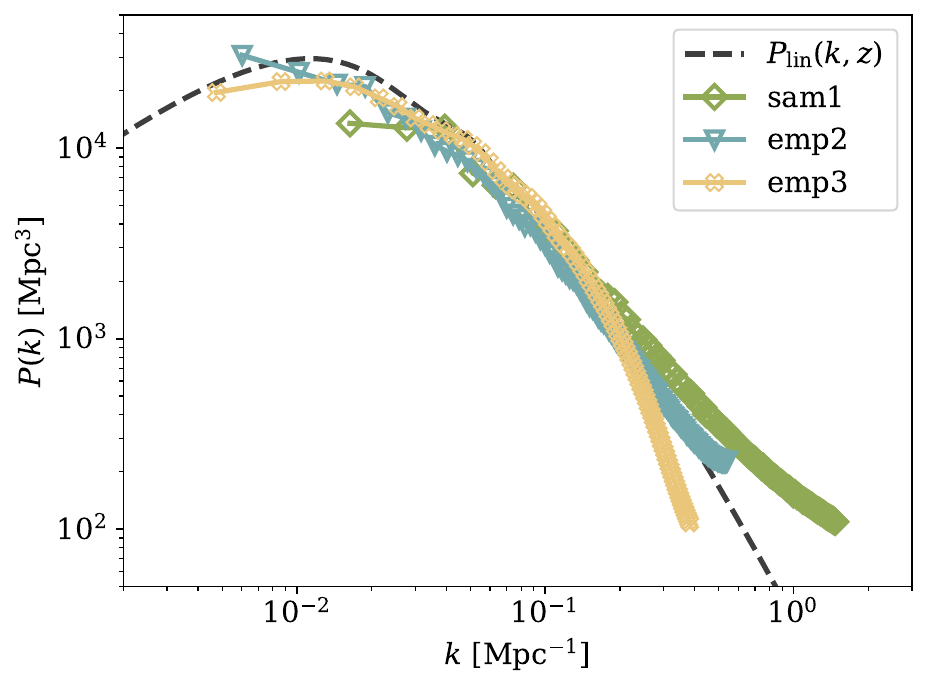}
    \caption{Marked power spectrum measured on the redshift $z=1$ boxes of sam1 (green diamonds connected by a green solid line), emp2 (blue downward triangles connected by blue solid line) and emp3 (orange crosses connected by the orange solid line). As a term of comparison, we over-plot as a black dashed line, the linear power spectrum of matter evolved linearly up to the same redshift of the simulations.}
    \label{fig:Pk}
\end{figure}
For our simulated samples, we define the mark as a function of the neutral hydrogen mass $M_{\mathrm{HI}}$ \citep[\eg][]{massara2021}.
We compare results from the $z=1$ snapshots of sam1, emp2 and emp3 \citep[note that, for sam1, measurements of $P_m(k)$ are also made available in][]{Spinelli_etal_2020}. 
In all cases\footnote{We have used estimators consistent with those available in the Pylians library (\href{https://pylians3.readthedocs.io/en/master/index.html}{pylians3.readthedocs.io}).} model results are shown only for $k_\text{min}<k<12 k_\text{Nyq}$, where $k_\text{min}$ is determined by the simulation size, while the $k_\text{Nyq}$ is the Nyquist frequency of the simulation, \ie the mean inter-tracer separation in real space. 
The black line in Fig.~\ref{fig:Pk} is the linear $P_\text{lin}(k,z=0)$ of matter at redshift $z=0$ evolved up to $z=1$ with the growth factor (i.e. multiplying by $[D(z=1)/D(z=0)]^2$). 
It has been obtained by running CAMB with approximated Planck2018 parameters \citep{Planck2018}.
The growth factor is computed with the \texttt{scampy.cosmology} module of the SCAMPy \citep{Ronconi2020} library, using rounded Planck2018 parameters (extracted from \texttt{astropy.cosmology.Planck2018} for consistency with the rest of the measurements).

All the methods show a slight loss in power for the largest scales (lowest $k$ values) with respect to the linearly evolved $P(k)$. 
The different values of $k_\text{min}$ here reflect the different sizes of the background DM simulations used by the authors.
In particular, emp2 and emp3 use approximated methods to build halo catalogues, that  allow lower values of $k_\text{min}$. 
On the other hand, sam1 computes the $P_m(k)$ from the galaxy distribution instead of haloes, which results in a higher value of $k_\text{Nyq}$. Besides this, galaxies are expected to be more affected by non-linear effects than dark matter haloes, as reflected by the excess power of the green diamonds at $k>0.1$ compared to both the empirical methods and the linear case.
The lower amplitude of $P_\text{HI}(k)$ measured from emp2 with respect to the linear (matter) case, together with the fact that this simulation is nearly complete in terms of the cosmological neutral hydrogen budget, is consistent with a low \HI bias, $b_\text{HI}\lesssim1$, i.e. \HI behaves as an almost unbiased tracer of the underlying dark-matter density field at these redshifts (see e.g.\ \citealt{Castorina2017, VillaescusaNavarro2018, Obuljen2023}).
On the other hand, the abrupt loss of power of emp3, at scales $k>0.2\ \text{Mpc}^{-1}$, is a numerical effect and is most likely due to aliasing generating from the coarser interpolation mesh used to perform the measurement.

Overall, the three methods analysed here show consistent results on scales $-2 \lesssim \log(k/{\rm Mpc}^{-1}) \lesssim -1$. 
At lower $k$-values, differences are mostly driven by cosmic variance, while at larger $k$-values the scatter is less straightforward to attribute to a single cause. 
Instead, it likely arises from a combination of factors, among which resolution and the nature of the modelled objects play a principal role.

\section{Context within the SKAO observational capabilities}\label{sec:SKAOcontext}

The models described in this Chapter are developed in the context of forthcoming observations with the SKAO, whose capabilities will open several complementary routes to probe neutral hydrogen across cosmic time. 
In the post-reionization regime, and especially for \textit{SKA-Mid}, simulated \HI maps and catalogues are not only a forecasting tool, but also the main bridge between physical models of how \HI populates haloes and galaxies and the survey observables that will be measured. 
This connection is particularly important because different observing strategies constrain different parts of the modelling, rather than all aspects equally well.

The telescopes of the SKAO, \textit{SKA-Mid} and \textit{SKA-Low}, probe complementary redshift ranges and physical regimes. \textit{SKA-Mid} is the key facility for the post-EoR applications discussed in this Chapter, since it will probe low to intermediate redshifts, roughly  $z \lesssim 3$, through both resolved \HI galaxy surveys and \HI intensity mapping. By contrast, \textit{SKA-Low} is aimed mainly at higher redshifts and the EoR, where semi-numerical methods \citep[such as 21cmFAST,][]{Mesinger2011,Murray2020} remain the natural tool for forecasting the large-scale 21-cm signal. In the \textit{SKA-Mid} context, the simulation methods reviewed here are especially well matched to survey design because they link the \HI content of galaxies and haloes to observable quantities such as number counts and clustering (see Sec.~\ref{sec:comparison}), and, with further post-processing brightness-temperature fluctuations and linewidths \citep[e.g.][]{mayor2026,nasirudin2026}.

Resolved \HI galaxy surveys will be particularly valuable for constraining the ingredients of semi-analytic and empirical models that govern the detectable galaxy population. 
The most directly testable components are the halo occupation of \HI galaxies (Sec.~\ref{sec:hihm}, available for all the models presented here), the redshift evolution of the \HI mass function (Sec.~\ref{sec:himf}, available for sam1, sam2 and emp1), and flux selection effects which vary model by model. 
The comparison among different modelling approaches detailed in this Chapter shows that predicted \HI galaxy number counts and large-scale bias can differ by factors of a few (Sec.~\ref{sec:pk}), and in some cases by nearly an order of magnitude at higher redshift (as evident from the comparison in Sec.~\ref{sec:omhi}), especially in the regime relevant to \textit{SKA-Mid} Band 1 surveys \citep[see also][]{nasirudin2026}. 
This means that future \HI galaxy observations with \textit{SKA-Mid} will be especially effective at testing how different models assign \HI to galaxies and haloes, and at identifying which prescriptions should be tuned to best reproduce the abundance and clustering of the detectable populations.

Intensity mapping (see \citet{Wolz01.2026.SKA} for a review), targeted by emp2 (Sec.~\ref{sec:method:hitz}) and emp3 (Sec.~\ref{sec:method:sinigaglia}) models directly, constrains a different layer of the same problem. 
On large scales, single-dish \HI intensity mapping is primarily sensitive to the combination of the cosmic \HI abundance and its large-scale bias, making it especially powerful for baryon acoustic oscillations, redshift-space distortions, and other large-volume cosmological measurements \citep{santos2015aska,Spinelli_etal_2020}. 
Because it captures the integrated emission from the full galaxy population, including sources far below the detection threshold, it provides a direct handle on the global \HI field rather than only on individually resolved systems. 
However, large-scale intensity mapping alone is less sensitive to the detailed internal distribution of \HI within haloes. 
Those aspects are more effectively constrained by interferometric observing mode, which reaches smaller scales and can probe the one-halo regime, where the signal depends on the \HI-halo mass relation, the radial profile of \HI within haloes, and the level of shot noise and halo-to-halo scatter \citep{Chen21, Spinelli_etal_2020,VillaescusaNavarro2018, Hitz2025}. 
In this sense, single-dish and interferometric intensity mapping play distinct but complementary roles: the former is best suited to large-scale cosmology, while the latter is more informative about the astrophysical structure of the \HI field itself.

Different telescope configurations and deployment phases (e.g. baseline and upgraded array setups) will impact the achievable sensitivity, angular resolution, and survey speed, and thus determine the range of scales and redshifts that can be effectively probed.
From the modelling side, however, the main lesson is that the approaches reviewed in this Chapter should be seen not simply as alternative simulation methods, but as a complementary toolkit whose components are constrained by different classes of SKAO observations.
Physically motivated models remain essential for connecting observables to the underlying baryonic processes that regulate the \HI content of galaxies and haloes \citep[see also][]{Lagos01.2026.SKA}, while empirical and fast approximate methods are crucial for producing the large ensembles of mock realizations needed for survey optimization, covariance estimation, and end-to-end pipeline validation \citep[e.g.][]{CosmoSWG2020,Hartley2023, Bonaldi2023}.

A key aspect of the interplay between simulations and observations is the forward modelling of instrumental effects.
Realistic \HI simulations, as those presented in this Chapter, can be processed through end-to-end pipelines that include beam convolution, noise realisations, survey masks, and foreground contamination, allowing direct comparison with expected observables such as angular and three-dimensional power spectra \citep[e.g.][]{Harrison01.2026.SKA,Harrison02.2026.SKA}.
This approach is essential for assessing the impact of systematics on cosmological measurements, including baryon acoustic oscillations, redshift-space distortions, and higher-order statistics.

The methodologies presented in this Chapter are highly needed to exploit the full potential of the SKAO, as demonstrated by the many Chapters in this book that make direct use of simulated data or of closely related techniques \citep[including:][]{Bertacca01.2026.SKA,Asorey01.2026.SKA,Harrison01.2026.SKA,Harrison02.2026.SKA,Wolz01.2026.SKA,Cunnington01.2026.SKA,Majumdar01.2026.SKA,Mayor01.2026.SKA,Nasirudin01.2026.SKA,Baker01.2026.SKA,Camera01.2026.SKA,Spinelli01.2026.SKA}.
The simulation strategies discussed here are broadly applicable across different array-assembly phases of the Observatory, and can be adapted to both AA$^{\star}$ and AA4. In practice, the main telescope specifications that affect the simulation design are the assumed survey area, depth, angular resolution, and limiting flux, many of which remain to be finalized at the time of writing. Some recent works \citep[e.g.][]{Ronconi2026} adopt survey assumptions motivated by proposed SKAO configurations \citep{CosmoSWG2020,Hartley2023}; however, such choices should still be regarded as indicative, since the final observing strategies will depend on the approved key science programmes and the capabilities available in the corresponding deployment phase.
By enabling robust predictions for \HI observables across different regimes, simulations play, and will continue to play, a central role in bridging theoretical modelling and observational analysis, both by enabling forecasts for a wide range of science cases and by supporting the interpretation of forthcoming data.
More generally, modelling the expected observables is one of the main tools through which the survey strategy itself can be optimized, by identifying which combinations of area, depth, and observing mode are best suited to maximize the scientific return of the Observatory.

\section{Discussion and future prospects}\label{sec:discussion}

The comparison presented in this Chapter highlights the diversity of existing approaches used to connect dark matter structures with the expected neutral hydrogen content in the Universe. 
The analysed methods differ significantly in their physical assumptions and calibration strategy.
Furthermore, while some methods provide information for individual galaxies, others only give halo-integrated quantities, making a one-to-one comparison more difficult for some observables. 

Overall, the various models provide qualitatively consistent correlations between the \HI and halo mass and a quite good agreement for integrated quantities, such as $\Omega_\text{HI}$ (Sec.~\ref{sec:omhi}) or the mean \HI bias (Sec.~\ref{sec:pk}), suggesting that large-scale statistics are relatively robust against the choice of modelling framework. At least part of this agreement is driven by the fact that models are tuned to reproduce the same observational data obtained in the local Universe. As a consequence, larger differences are found when considering model prediction at higher redshifts where observational constraints are more scarce or sometimes based on different tracers (as is the case for the cosmic density of \HI). 

Systematic differences emerge when comparing the detailed shape and scatter of the HIHM relation or the evolution of the \HI distribution with redshift. 
Part of these differences can be traced back to the intrinsic nature of the models. 
Semi-analytical approaches, such as sam1 (Sec.~\ref{sec:method:GAEA}) and sam2 (Sec.~\ref{sec:method:yates}), self-consistently follow baryonic processes and remain sensitive to the adopted prescriptions. 
Empirical methods, on the other hand, are computationally efficient and directly calibrated on observations (as in emp1 from Sec.~\ref{sec:method:ronconi} and emp2 from Sec.~\ref{sec:method:hitz}) or hydrodynamical predictions (as in emp3 from Sec.~\ref{sec:method:sinigaglia}).

In general, any behaviour in regimes (in terms of e.g. halo mass or redshift) beyond the calibration range considered, are extrapolations that should be considered with caution. 
The different nature of the models, and the weaknesses and advantages briefly discussed, naturally lead to a different tool selection when addressing specific questions/problems. 
Clearly, when developing specific algorithms for data analysis, empirical methods provide an efficient and natural framework for testing in realistic cases, at least over the redshift range where observational data are available to construct the models. 
However, it is difficult to translate the information obtained employing statistical methods into constraints on physical processes driving baryonic evolution unless explicitly accounting for them, like done for example in semi-analytic models. As mentioned above, the observational information currently available beyond $z=0$ is scarce but the situation is rapidly evolving with ongoing survey efforts and will be ultimately revolutionized as the SKAO will move into operation. The tools we have discussed will play a crucial role for the exploitation of the huge amount of information that will be collected in the coming years. 

Beyond the analysis presented here, all methods described in this work can be post-processed to derive the corresponding \HI brightness temperature fields enabling additional applications, for example, in the framework of  intensity mapping studies \citep[\eg][]{Spinelli01.2026.SKA}, by calculating the 21cm emission lines \citep{Mayor01.2026.SKA}.
Furthermore, at least some of the methods discussed in this Chapter, also predict many other additional properties, beyond masses and positions.
For example, both the semi-analytic models (Sec.~\ref{sec:methods:sam}) provide information on AGN feedback, gas partitioning and evolutionary properties, useful for understanding, e.g., the formation processes of the sources modelled.
The HIHM-relation-based model (Sec.~\ref{sec:method:sinigaglia}), provides environmental information and, the T-RECS-based method (Sec.~\ref{sec:method:ronconi}), also includes emission properties for other populations of sources, providing a natural framework for, e.g., cross-correlation studies.

The diversity in outcomes, far from being a limitation, illustrates the sensitivity of \HI modelling to both physical assumptions and methodological choices.
Together, these results emphasise the importance of maintaining a heterogeneous modelling ecosystem. 
Each approach captures complementary aspects of the complex connection between dark matter and baryons.
Different scientific problems call for different modelling strategies rather than a single "best" prescription: one algorithm may excel for halo-scale \HI assignments, another for large-scale intensity mapping, and another for hybrid clustering or cross-correlation forecasts. 
This Chapter’s objective has been to illustrate this methodological diversity and to inform the reader about the regimes in which each model is likely to perform more reliably. 

Looking ahead, the new generation of \HI surveys, including ASKAP’s WALLABY \citep[mapping up to $z\sim0.1$ over $\sim14\,000\,\mathrm{deg}^2$ with $\sim30''$ resolution,][]{wallaby2020}, and deep surveys with MeerKAT such as MIGHTEE-HI and LADUMA \citep[probing to $z\sim0.3$ and $z\sim1.4$, respectively,][]{mightee2021, laduma2016}, will yield independent observational constraints on the neutral hydrogen distribution. 
These upcoming data will allow modelling paradigms, such as those presented in the previous Sections, to be recalibrated and validated in preparation of the SKAO era. 
By benchmarking models against real survey outputs, it will be possible to gradually refine parameter choices, to reduce systematic uncertainties, and to build a more robust interpretative framework for SKAO’s unprecedented \HI survey capabilities.

\section*{Author List Ordering}

The authors of this Chapter are grouped into three tiers reflecting their  contribution. The first tier, ordered by contribution, includes TR, GDL, and MS, who led the conceptual development and writing of this work. The second tier (FF, PH, FS, LX, RMY), ordered alphabetically, includes authors who contributed actively through the production and analysis of the simulation results presented. The third tier, also ordered alphabetically, includes authors who participated in the scientific discussion, interpretation of results, and revision of the manuscript.

\section*{Acknowledgements}

PH is partially supported by grant CRSII5\_193826 from the Swiss National Science Foundation, and by SERI as part of the SKACH consortium. JLB acknowledges funding from the grant UC-LIME (PID2022-140670NA-I00), financed by MCIN/AEI/ 10.13039/501100011033/FEDER, UE.

%%%%%%%%%%%%%%%%%%%%%%%%%%%%%%%%%%%%%%%%%%%%%%%%%%%%%%%%%%%%%%%%%%%%%%%

\newpage
\bibliographystyle{abbrvnat-maxbibnames4}
\bibliography{chapter,robyates} % if your bibtex file is called chapter.bib

%%%%%%%%%%%%%%%%%%%%%%%%%%%%%%%%%%%%%%%%%%%%%%%%%%%%%%%%%%%%%%%%%%%%%%%

\end{document}

%% file: tab/cosmologies.tex
\begin{tabular}{lccccccr}
    \hline
    DM Simulation & $h$ & $\Omega_\text{CDM}$ & $\Omega_\Lambda$ & $\Omega_b$ & $\sigma_8$ & $n_s$ & Reference \\
    \hline
    \multicolumn{8}{c}{Semi-Analytic Models}\\
    \hline
    &&&&&&\\ %insert your method's parameters here
    \multicolumn{8}{c}{\textbf{sam1:} \textit{GAEA} (Sec.~\ref{sec:method:GAEA})}\\
    %Millennium & 0.73 & 0.25 & 0.75 & 0.0455 & 0.9 & 1 & \\
    %Millennium-II & 0.73 & 0.25 & 0.75 & 0.0455 & 0.9 & 1 & \\
    P-Millennium & 0.6777 & 0.307 & 0.693 & 0.04825 & 0.8288 & 0.9611 & (1)\\
    &&&&&&\\
    \multicolumn{8}{c}{\textbf{sam2:} \textit{L-Galaxies} (Sec.~\ref{sec:method:yates})}\\
    Millennium & 0.673 & 0.2663 & 0.685 & 0.0487 & 0.829 & 0.96 & (2)\\
    Millennium-II & 0.673 & 0.2663 & 0.685 & 0.0487 & 0.829 & 0.96 & (3)\\
    %FLARES-DMO & 0.6777 & 0.307 & 0.693 & 0.04825 & 0.8288 & 0.9611 & \\
    &&&&&&\\
    \hline
    \multicolumn{8}{c}{Empirical Models}\\
    \hline
    &&&&&&\\ %insert your method's parameters here
    \multicolumn{8}{c}{\textbf{emp1:} \textit{TRECS+SCAM} (Sec.~\ref{sec:method:ronconi})}\\
    DEMNUni-HR & 0.670 & 0.320 & 0.680 & 0.050 & 0.834 & 0.962 & \eg (4, 5, 6)\\
    &&&&&&\\
    \multicolumn{8}{c}{\textbf{emp2:} \textit{HI Halo Model} (Sec.~\ref{sec:method:hitz})}\\
    PINOCCHIO & 0.6736 & 0.3153 & 0.6847 & 0.0493 & 0.8111 & 0.9649 & (7)\\
    &&&&&&\\
    \multicolumn{8}{c}{\textbf{emp3:} \textit{HIHM relation} (Sec.~\ref{sec:method:sinigaglia})}\\
    ALPT & 0.6777 & 0.307 & 0.693 & 0.04825 & 0.8288 & 0.9611 & (8)\\
    &&&&&&\\
    \hline
\end{tabular}